\documentclass[pof,aps,twocolumn]{revtex4-1}
\usepackage{graphicx} 
\usepackage[usenames]{color}
\usepackage{amsmath,amssymb}
\usepackage{gensymb}
\usepackage{natbib}
\usepackage{color}
\def\strutdepth{\dp\strutbox}
\def\nw#1{\strut\vadjust{\kern-\strutdepth\vtop to0pt{\vss\hbox to\hsize
{\hskip\hsize\hskip5pt$\leftarrow$\hss\strut}}}{\em #1}}

\begin{document}

\title{Elastic deformation due to tangential capillary forces}
\author{Siddhartha Das$^1$, Antonin Marchand$^2$, Bruno Andreotti$^2$ and Jacco H. Snoeijer$^1$.}
\affiliation{
$^{1}$Physics of Fluids Group and J. M. Burgers Centre for Fluid Dynamics,
University of Twente, P.O. Box 217, 7500 AE Enschede, The Netherlands\\
$^{2}$Physique et M\'ecanique des Milieux H\'et\'erog\`enes, UMR
7636 ESPCI -CNRS, Univ. Paris-Diderot, 10 rue Vauquelin, 75005, Paris.
}

\begin{abstract}
A sessile liquid drop can deform the substrate on which it rests if the solid is sufficiently ``soft". In this paper we compute the detailed spatial structure of the capillary forces exerted by the drop on the solid substrate using a model based on Density Functional Theory. We show that, in addition to the normal forces, the drop exerts a previously unaccounted tangential force. The resultant effect on the solid is a pulling force near the contact line directed towards the interior of the drop, i.e. not along the interface. The resulting elastic deformations of the solid are worked out and illustrate the importance of the tangential forces.    
\end{abstract}

\maketitle

\section{Introduction}
\begin{figure}[t!]
\begin{center}
\includegraphics{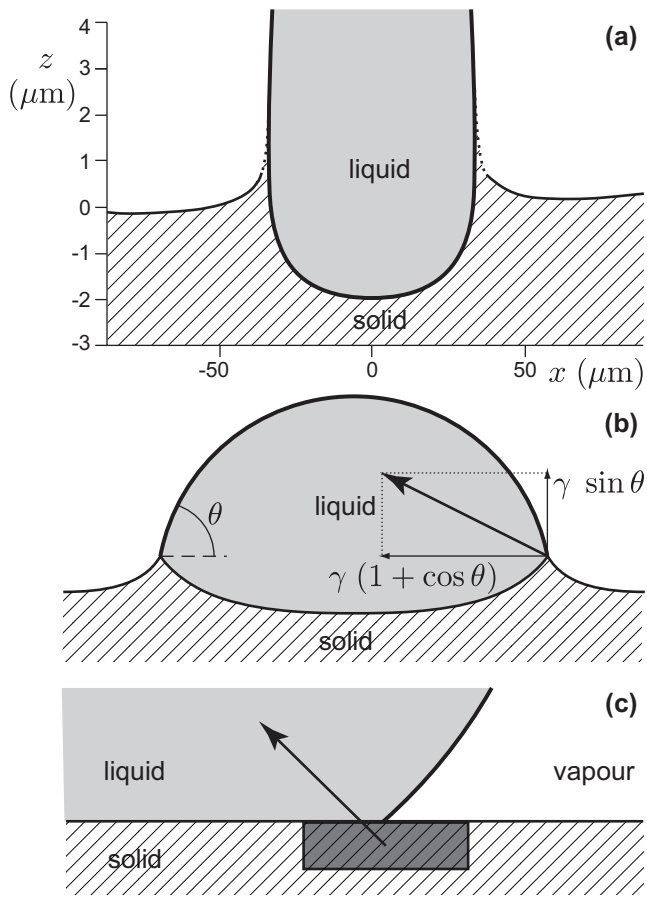}
\end{center}
\caption{(a) Shape of an ionic liquid drop on a soft PDMS surface, determined experimentally by Pericet-C'amara et. al. \cite{PericetLANGMUIR} using laser scanning microscopy. The aspect ratio has been exaggerated by a factor $15$. (b) Macroscopic representation of the capillary forces below a small liquid drop and resulting elastic displacements. A localized normal force of magnitude $\gamma \sin \theta$ pulls along the contact line and is balanced by the Laplace pressure, which is applied to the wetted surface area. In this paper we show that there is also a nonzero tangential force of magnitude $\gamma (1+\cos \theta)$. The resulting force points towards the interior of the liquid, at an angle $\alpha$ to the horizontal. The sum of these forces induces a small deformation of the interface. (c) Physical origin of the tangential force. Solid molecules just below the contact line experience a resultant attraction towards the interior of the drop, where there is more liquid. This effect is particularly striking when the contact angle $\theta$ is above $\pi/2$.}
\label{fig:sketch}
\end{figure}

A drop of liquid can adhere to a solid surface due to molecular interactions. The microscopic structure of the forces exerted on the liquid is accounted for by classical wetting theory, e.g., using the concept of disjoining pressure~\cite{deGennesRMP,BonnRMP}. From a macroscopic perspective this induces a normal force localized near the contact line of magnitude $\gamma \sin \theta$~\cite{deGennes,WhiteJCIS}, where $\gamma$ is the liquid-vapor surface tension and $\theta$ the contact angle. This attractive force (per unit line) is compensated for by a repulsive Laplace pressure spread out over the contact area of the drop, giving a zero resultant force. Experimental evidence for such nontrivial spatial structure of capillary forces can be inferred from the elastic deformation of the solid below a liquid drop~\cite{RusanovUSSR,LesterJCIS,WhiteJCIS,YukJCIS,PericetLANGMUIR,YuJCIS,CamaraCPC,CamaraSM,PyEPJST}. Figure~\ref{fig:sketch}a (taken from \cite{PericetLANGMUIR}), shows that elastic deformation is small, but measurable, on sufficiently soft substrates. 

One should note, however, that elastic deformation is induced by forces exerted \emph{on the solid}. This means that, \emph{a priori}, one cannot apply the classical theories of wetting, which account for the forces exerted \emph{on the liquid}. The goal of the present paper is therefore to extend the standard formalism of wetting to compute the microscopic forces exerted on the solid. It will turn out that, indeed, there are some crucial differences when changing the ``system" to which the forces are applied from liquid to solid. We will work out the consequences for the resulting elastic deformations, which we show to be fundamentally different from existing theoretical predictions.

There has been an intrinsic interest in the elasto-capillary interactions such as drops on a soft surface~\cite{PyEPJST, PyPRL, PinSM, BicoJPCM,HonschotenAPL,JerisonPRL}. For example, the contact angle of the drop on a soft surface is slightly different from Young's equilibrium angle $\theta$~\cite{FortesJCIS, KernSS, OlivesJPCM, OlivesJPCM1}. The energy balance now has an elastic contribution and for a small deviation from Young's angle $\Delta\theta$ one finds $ \Delta\theta \sim \gamma/(ER)$, where $E$ is the Young elastic modulus of the system and $R$ the radius of the drop. Other effects are enhancement of contact angle hysteresis \cite{ExtrandJCIS}, visco-elastic dissipation during drop spreading~\cite{ShanahanLANGMUIR, CarreJCIS} or enhanced nucleation density in condensation processes~\cite{SokulerLANGMUIR}. To estimate the significance of elastic deformations it is instructive to consider the characteristic length scale $\ell=\gamma/E$, from the ratio of surface tension and the elastic modulus, which determines the scale of deformations. For polycrystalline materials, for which the elastic modulus is determined by the interaction between molecules ($E\sim $ GPa), such elastic deformations are irrelevant ($\ell \sim 10^{-12}$~m). However, for solids whose elasticity is of entropic origin (rubbers or gels with $E \lesssim 10$~kPa) deformations extend into the micron range and have indeed been observed experimentally~\cite{CarreNATURE, CamaraSM, PericetLANGMUIR, YuJCIS, PyEPJST,JerisonPRL}. 

Interestingly, the elastic problem cannot be solved within a macroscopic framework where the effect of surface tension is represented by a localized force. This is because the elastic response to a Dirac $\delta$-function gives an infinite displacement at the contact line. This singularity can be regularized by introducing microscopic physics, either in the solid phase \cite{DaiPRE} or in the liquid phase. The simplest and most common approach is to spread out the pull of the contact line over a small region of finite thickness $w$. By assuming that this force is uniformly distributed in this thin region, Rusanov~\cite{RusanovUSSR} and Lester~\cite{LesterJCIS} have solved the elastic displacement profile analytically. The resulting scale of displacement near the contact line becomes $\sim \ell\,\ln(R/w)$, illustrating the singularity as $w \rightarrow 0$. As pointed out by White~\cite{WhiteJCIS}, the precise distribution of capillary forces can be obtained explicitly when introducing a microscopic model for the molecular interactions. It was argued that the normal traction is simply equal to the disjoining pressure $\Pi(h)$. This indeed gives a pressure peak localized near the contact line, whose integral is equal to $\gamma \sin \theta$. Once more it should be emphasized, however, that this involves the forces exerted on the liquid, and not the forces exerted on the solid, as required to obtain the elastic deformation.

In this paper we compute the forces exerted by the liquid on the solid using Density Functional Theory in the sharp-kink approximation~\cite{GettaPRE}. This approach has the merit that (for small contact angles) all results can be expressed in terms of the disjoining pressure and can thus be related to the usual wetting theory. Our central finding is that, besides a normal component, there is also a nonzero tangential component exerted on the solid, as depicted in Fig.~\ref{fig:sketch}b. The physical origin of this tangential force can be understood by considering solid molecules directly below the contact line: there is always more liquid present at the interior of the contact line than at the exterior, yielding an ``inward" attractive interaction with a nonzero tangential component. Near the contact line, this leads to the following normal ($F_n$) and tangential ($F_t$) force components: 
\begin{eqnarray}
F_n/L &=& \gamma \sin\theta, \label{eq:Fn}\\
F_t/L &=& \gamma (1+\cos\theta) \label{eq:Ft},
\end{eqnarray}
where $L$ is the length of the contact line. Hence, the resultant force points towards the liquid (Fig.~\ref{fig:sketch}c) and is not directed along the interface. The angle that characterizes the direction of the force follows from $\tan{\alpha}=\sin{\theta}/(1+\cos{\theta})$, which simply gives $\alpha=\theta/2$. This force was not considered before in the literature and we therefore investigate the way it affects the elastic deformation of the substrate.

It is worth emphasizing that the capillary forces predicted by the Density Functional Theory model are perfectly consistent with thermodynamics and do not lead to a violation of YoungÕs law. One should bear in mind that (1,2) represent the forces exerted on the solid Ð by contrast, YoungÕs law reflects the equilibrium shape of the deformable liquid and thus involves forces exerted on the liquid \cite{MarchandAJP}. In fact, the same model was previously shown to yield YoungÕs angle when computing the equilibrium shape of the liquid \cite{MerchantPFA, SnoeijerPOF} and thus yields a consistent thermodynamic picture. Let us further note that the presented results are fundamentally different from the tangential contribution described in Refs.~\cite{PericetLANGMUIR, RusanovUSSR, YukJCIS}. Those refer to forces originating from the difference with Young's angle induced by deformation itself. This is a higher order effect with respect to the basic tangential capillary forces and are smaller by orders of magnitude ($\sim\ell/R$). 

The paper is organized as follows. Section~\ref{sec:forces} develops the formalism used to calculate the forces on the solid, and points out the relation with disjoining pressure and thermodynamic pressure. In Sec.~\ref{sec:results} we then quantify the tangential and normal forces, leading to (\ref{eq:Fn},\ref{eq:Ft}), and illustrate the results for a Van der Waals model for the interactions in Sec.~\ref{sec:numerics}. In Sec.~\ref{sec:elastic} we then show how the tangential capillary force affects the elastic deformation of the solid in the limit of small deformations, and the paper closes with a discussion. 

\section{Density Functional Theory in the sharp-kink approximation}\label{sec:forces}

Here we introduce the formalism needed to compute the forces exerted by a liquid drop on a solid. This is based on Density Functional Theory in the sharp-kink approximation, in the same spirit as~\cite{MerchantPFA,GettaPRE,SnoeijerPOF,BauerEPJB,WeijsPF}. We will define four important pressures: the disjoining pressure $\Pi$, the thermodynamic pressure $P_L$, the repulsion $p_r$ and the Laplace pressure. These will be used in Sec.~\ref{sec:results} to make explicit predictions for the capillary forces on the solid.

\subsection{Repulsion $p_r$}
From standard continuum mechanics one can express the force $\mathbf{F}$ that the liquid exerts on the solid as
\begin{equation}\label{eq:force}
\mathbf{F}= - \int_{\cal S} \mathbf{dr} \nabla \phi_{LS} 
- \int_{\partial \mathcal{S}} \mathbf{dA} \,p_r,
\end{equation}
where ${\cal S}$ and $\partial \mathcal{S}$ are the volume and boundary of the solid domain respectively. Long-range (attractive) interaction between liquid and solid molecules gives rise to a potential $\phi_{LS}$ inside the solid i.e. an energy per unit volume. The short-range (repulsive) interactions are assumed to lead to isotropic stresses and to determine completely the pair correlation functions at a given density. The liquid therefore exerts a contact pressure $p_r$ on the solid. Such an explicit separation in long-range attraction and short-range repulsion is a standard approach in Density Functional Theory~\cite{HansenB}, and forms the basis for our analysis. We furthermore use the so-called sharp interface approximation~\cite{MerchantPFA,GettaPRE,BauerEPJB}, where the different phases are assumed to be homogeneous and separated by an infinitely thin interface. In this framework, one can express the liquid-on-solid potential as
\begin{equation}
\phi_{LS}(\mathbf{r} ) = \rho_L \rho_S \int_{\cal L} \mathbf{dr'} \varphi_{LS}(|\mathbf{r} -\mathbf{r'}|),
\end{equation}
where $\varphi_{LS}$ is the effective solid-liquid interaction potential, which takes into account the pair correlation function (throughout we assume the interaction to depend only on the intermolecular distance $|\mathbf{r} -\mathbf{r'}|$). $\rho_L$ and $\rho_S$ are the liquid and solid densities respectively. The integral runs over the entire liquid domain ${\cal L}$, which implies that $\phi_{LS}$ is a functional of ${\cal L}$. More generally, one can define 
\begin{equation}
\phi_{\alpha \beta}(\mathbf{r} ) = \rho_\alpha \rho_\beta \int_{\cal \alpha} \mathbf{dr'}\varphi_{\alpha \beta}(|\mathbf{r} -\mathbf{r'}| ),
\end{equation}
as the potential in phase $\beta$ due to phase $\alpha$. Note that in general $\phi_{\alpha \beta}\neq \phi_{\beta \alpha}$, unless the domains of $\alpha$ and $\beta$ have an identical shape.

The calculation of the force $\mathbf{F}$ (or to be precise the normal component $F_n$), still requires the repulsive pressure $p_r$. This can be obtained from the equilibrium condition inside the liquid,
\begin{equation}
\label{eq:liquid}
\nabla \left( p_r+\phi_{LL}+\phi_{SL} \right) = 0
\end{equation}
so that the total potential $p_r+\phi_{LL}+\phi_{SL}$ must be homogeneous \cite{GettaPRE, MerchantPFA}.  Here it is assumed that the incompressibility of the liquid results from the repulsive term $p_r$, which adapts its value according to the local values of the liquid-liquid and solid-liquid interactions $\phi_{LL},\phi_{SL}$. 

\subsection{Thermodynamic pressure $P_L$}

The existence of a quantity conserved throughout the liquid (the total potential) allows to generalize the concept of thermodynamic pressure $P_L$ of the liquid phase.  Let us consider the case where the liquid is in equilibrium with its vapour phase. The constant of integration in (\ref{eq:liquid}) can then be determined from the boundary conditions at the interfaces. In the present framework repulsion is described as a contact force, as opposed to attraction whose influence is spread overs a few molecular diameters. As a consequence the repulsive potential $p_r$ must be continuous, while there is no such condition for $\phi_{LL}$ and $\phi_{SL}$. Inside the vapour the density is so small that interactions can be neglected compared to the kinetic pressure $P_v$. Therefore, the only force transmission by the vapor on the liquid is through $p_r$ and we simply find $p_r=P_v$ in the vapor. When the liquid-vapor interface is flat, we thus find $p_r = P_v=P_L$ at the both sides of the interface. If in addition the liquid film is macroscopically thick (meaning that $\phi_{SL}=0$ at the interface), we thus obtain inside the liquid
\begin{equation}\label{eq:P}
p_r+\phi_{LL}+\phi_{SL} = P_L + \Pi_{LL}(0),
\end{equation}
where we denote $\Pi_{LL}(0)$ as the value of  $\phi_{LL}$ at the edge of an infinite half space of liquid. For a detailed definition of $\Pi_{\alpha\beta}$ we refer to the subsequent paragraphs. 

We propose to take (\ref{eq:P}) as a definition of the thermodynamic pressure $P_L$ in the liquid. It is, as expected,  a characteristic of the phase and it can be identified to the true thermodynamic pressure in the macroscopic limit.

\subsection{Disjoining pressure $\Pi$}\label{subsec:Disj_Pres}

\begin{figure}[t!]
\includegraphics{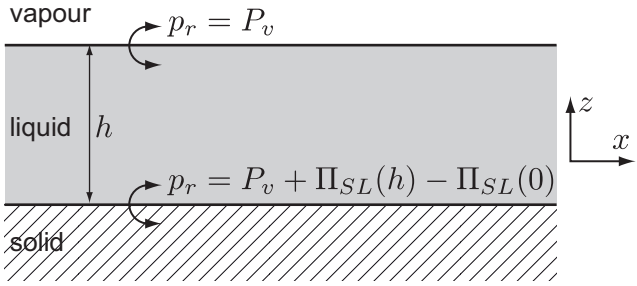}
\caption{Repulsion $p_r$ for a flat liquid film of thickness $h$. The potentials $\Pi_{LL}$ and $\Pi_{SL}$ are defined in the text.}
\label{FlatInterface}
\end{figure}

The disjoining pressure is defined as the pressure acting on a flat liquid film of thickness $h$, due to the influence of the solid substrate (Fig.~\ref{FlatInterface}). It can be related to the potential $\Pi_{\alpha \beta}(h)$ at a distance $h$ from an infinite half space: 
\begin{eqnarray}
\Pi_{\alpha \beta}(h) &=& \int_{-\infty}^\infty dx\int_{-\infty}^\infty dy \int_h^\infty dz \, \varphi_{\alpha \beta}(|\mathbf{r}|)\nonumber\\
&=&2\pi\int_{h}^\infty r^2\left(1-\frac hr\right)\varphi_{\alpha \beta}(r)dr\;.
\label{eq:gen_Pi_phi_rel}
\end{eqnarray}
Contrarily to $\phi_{\alpha\beta}\ne\phi_{\beta\alpha}$, we recover the symmetry property $\Pi_{\alpha\beta}=\Pi_{\beta\alpha}$.
Within the sharp-kink approximation discussed above, the disjoining pressure $\Pi(h)$ can then be computed as~\cite{SnoeijerPOF,BauerEPJB}
\begin{equation}
\Pi(h) = \Pi_{SL}(h)- \Pi_{LL}(h).
\label{Disjoining_Pressure}
\end{equation}
From this definition it is clear that $\Pi(h)$ quantifies the change in energy when replacing part of the liquid by solid molecules. In other words, it is the correction to the total potential at the liquid vapour interface due to the presence of the solid at a distance $h$.

To relate to the macroscopic concepts of surface tensions and contact angles, one can use the following integral properties~\cite{SnoeijerPOF}:
\begin{eqnarray}
\int_{0}^\infty \Pi_{LL}\left(h\right)dh&=&-2\gamma, \label{int_phi_LL} \\
\int_{0}^\infty\Pi_{SL}\left(h\right)dh &=&-\gamma_{SV}+\gamma_{SL}-\gamma  \nonumber \\
&=& -\gamma\left(1+\cos \theta \right)\;. \label{int_phi_SL}
\end{eqnarray}
Note that the solid-liquid interaction only involves the difference $\gamma_{SL}-\gamma_{SV}$, and not the surface tensions individually.
Here we made use of Young's law, which was explicitly validated in the context of the present DFT model \cite{MerchantPFA,SnoeijerPOF}.
Combining the two equations, one obtains the usual normalization of the disjoining pressure:
\begin{eqnarray}
\int_{0}^\infty\Pi\left(h\right)dh&=&-\gamma_{SV}+\gamma_{SL}+\gamma \nonumber \\
&=& \gamma\left(1-\cos \theta \right)\;.
\label{int_phi_Disj_h}
\end{eqnarray}
Note that, in the case of a description that includes a precursor film of thickness $h_0$, the lower limit of integration should be $h_0$ instead of $0$ for each of the integrals.

\begin{figure}[t!]
\includegraphics{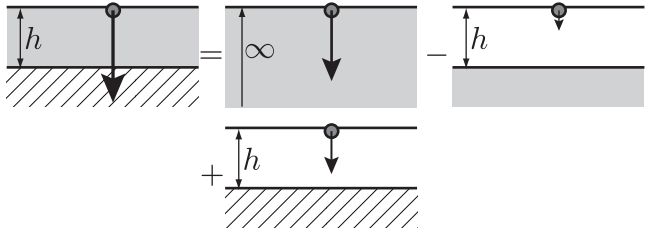}
\caption{The potential energy at the top surface of a liquid film of thickness $h$ can be decomposed into three contributions. The solid-liquid interaction $\phi_{SL}$ is simply the potential due to an infinite half-space of solid located at distance $h$. The liquid-liquid interaction $\phi_{LL}$ can be expressed into two contributions of semi-infinite liquid phases: in one case it is located at a distance $h$, while in other case it is at a zero distance.}
\label{DecompositionA}
\end{figure}

We can now provide the detailed structure of the potential fields inside the flat liquid film (Fig.~\ref{FlatInterface}). Once again, the density inside the vapor can be neglected as compared to the kinetic pressure $P_v$ and we simply find $p_r=P_v$. Crossing the liquid vapour interface, the repulsive pressure $p_r$ is continuous. The attractive potential at the liquid-vapour interface can be decomposed as shown in Fig.~\ref{DecompositionA}. The attraction by the solid leads to a potential $\phi_{SL}=\Pi_{SL}\left(h\right)$. Namely, the solid represents an infinite half-space at a distance $h$. The self-attraction by the liquid film can be also expressed as a function of the interaction induced by half a space of liquid located at a distance $h$: from the geometric construction in Fig.~\ref{DecompositionA} one finds $\phi_{LL}=\Pi_{LL}\left(0\right)-\Pi_{LL}\left(h\right)$. We finally get the expression of the thermodynamic pressure from Eq.~(\ref{eq:P}):
\begin{eqnarray}
P_L &=&P_v+ \Pi_{SL}(h) - \Pi_{LL}(h) \nonumber \\
&=& P_v+\Pi(h),
\label{Disjoining_Pressure1}
\end{eqnarray}
where we used the definition of the disjoining pressure (\ref{Disjoining_Pressure}). For a thin flat film, the thermodynamic pressure inside the liquid is thus different from the vapour pressure. The difference is precisely the disjoining pressure. This justifies and explains the concept.

\subsection{Laplace pressure}

\begin{figure}[t]
\begin{center}
\includegraphics{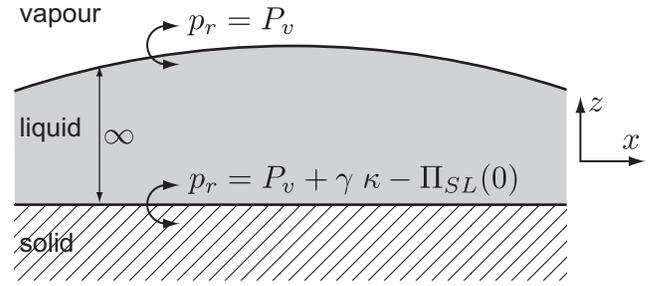}
\caption{\label{CurvedInterface}Repulsion $p_r$ for a macroscopic film of liquid that has a curved liquid-vapour interface. By convention, the curvature $\kappa$ is positive on the schematic.}
\end{center}
\end{figure}

When the interfaces are curved, it can be shown by expansion of the domain of integration \cite{MerchantPFA,GettaPRE} that there is a deficit of energy that is exactly equal to the Laplace pressure. Consider the liquid film shown in Fig.~\ref{CurvedInterface}, which is assumed to be much thicker than the molecular size. The liquid-vapor interface  presents a curvature $\kappa$. If the curvature is macroscopic, $h_0 |\kappa| \ll 1$, the potential due to the liquid film turns out to be
\begin{equation}
\label{laplaceLL}
\phi_{LL}=\Pi_{LL}(0)+\gamma\;\kappa\;.
\end{equation}
If $\kappa>0$, the liquid domain is convex so that the volume of liquid attracting a given element of liquid at the surface is smaller than in the flat case. Therefore, $|\phi_{LL}|<|\Pi_{LL}(0)|$ --~note that $\phi_{LL}$ and $\Pi_{LL}$ are negative. We finally get the expression of the thermodynamic pressure from Eq.~(\ref{eq:P}):
\begin{equation}
P_L =P_v+\gamma\;\kappa.
\label{Laplace_Pressure}
\end{equation}
Indeed, one recognises the standard Laplace pressure. We see here that, just like the disjoining pressure, the Laplace pressure results from the geometry of the attractive volume. 

\subsection{Local approximation}

In principle the above equations are sufficient to evaluate the liquid-on-solid force $\mathbf{F}$, once the molecular interaction $\varphi_{LS}$ is specified. However, the technical difficulty is that both $\phi_{LS}$ and $p_r$ are functionals of the liquid domain ${\cal L}$. In turn, the shape of this domain has to be found self-consistently such that the liquid is in equilibrium~\cite{SnoeijerPOF,GettaPRE,BauerEPJB,MerchantPFA,WeijsPF} i.e. by minimizing the total free energy. For a liquid-vapor interface, the equilibrium condition is that the interface is an iso-potential. As $p_r=P_v$ is constant, this imposes that $\phi_{LL}+\phi_{SL}$ is also constant along the free surface. This can readily be inferred from Eq.~(\ref{eq:P}). It was shown in \cite{MerchantPFA,SnoeijerPOF} that this framework indeed leads to YoungÕs law for the macroscopic contact angle of the liquid at equilibrium.

Th equilibrium condition can be simplified for interfaces of macroscopic curvatures and small slopes~\cite{SnoeijerPOF,BauerEPJB}. Then, the nonlocal \emph{functionals} reduce to \emph{functions} of the local thickness $h$ and local curvature $\kappa$. The two geometrical effects previously described add to each other: the capillary pressure is the sum of the Laplace pressure and the disjoining pressure. At the free surface, the solid induced potential reads $\phi_{SL}= \Pi_{SL}(h)$ and the liquid induced potential reads:
\begin{equation}
\phi_{LL} = \Pi_{LL}(0) - \Pi_{LL}(h) + \gamma \kappa~,
\end{equation}
In fact, one can show from a geometric construction that this local approximation is valid when $h \gg h_0 h'$, which for small slopes spans the entire liquid domain. 

One finally obtains the thermodynamic pressure $P_L$ from Eq.~(\ref{eq:P}): 
\begin{eqnarray}
P_L &=& P_v -\Pi_{LL}(h) + \Pi_{SL}(h) + \gamma \kappa \nonumber \\
&=& P_v+ \gamma \kappa + \Pi(h)~.\label{eq:dropeq}
\end{eqnarray}
This equation can be used to compute the equilibrium shape of the liquid domain, expressed as $h(x,y)$. This will be done explicitly in Sec.~\ref{sec:numerics}, where we compute the shape of drops and capillary forces for a specific choice of $\Pi(h)$.

\section{Capillary forces on a solid}\label{sec:results}

We will now explicitly compute the force exerted by the liquid on the solid, $\mathbf{F}$, starting from (\ref{eq:force}). The Density Functional model discussed above will allow for direct evaluation of $\mathbf{F}$, and also provides the detailed spatial distribution of the capillary forces in the vicinity of a contact line. Throughout this section we assume that the solid is undeformed.

\subsection{Capillary forces under a macroscopic film}\label{sec:laplace}

Far from a contact line,  when the liquid film is macroscopic, one expects the force per unit area acting on the solid to be simply the thermodynamic pressure $P_L$, transmitted at the interface. Here we check that this property is verified by the Density Functional Theory model.

We consider again a perfectly flat solid-liquid interface separating an infinite half space of solid from a macroscopic drop of liquid (Fig.~\ref{CurvedInterface}). The only gradient of $\phi_{LS}$ then arises in the direction normal to the interface. Therefore, Eq. (\ref{eq:force}) can be integrated along the normal direction yielding a normal force inside the solid,
\begin{equation}
F_n =- \int_{\partial \mathcal{S}} \mathbf{dA} \left[ \phi_{LS} + p_r \right], \label{eq:normal} 
\end{equation}
where the integrand of (\ref{eq:normal}) has to be evaluated at the solid-liquid interface. Since the volume of liquid ${\cal L}$ is almost a semi-infinite domain bounded by a flat interface, we can write $\phi_{LS} \simeq \Pi_{LS}(0)=\Pi_{SL}(0)$ and $\phi_{LL} \simeq \Pi_{LL}(0)$.  Using Eq.~(\ref{eq:P}), we obtain $p_r=P_L-\Pi_{SL}(0)$. Finally, this yields the total normal stress acting on the solid (Fig.~\ref{CurvedInterface}):
\begin{eqnarray}
f_n&=&-p_r-\Pi_{LS}(0)=-P_L \nonumber \\
&=&-P_v-\gamma\kappa,
\label{stresslaplace}
\end{eqnarray}
which is the force per unit area. For a convex liquid interface ($\kappa>0$), the solid is submitted to a higher (Laplace) pressure.

\begin{figure}[t!]
\includegraphics{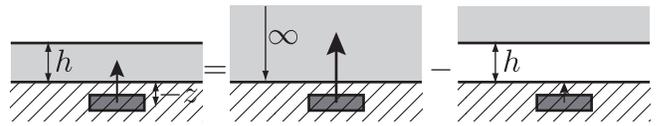}
\caption{Interaction of a volume element within the solid (at position $z<0$) with the liquid film of thickness $h$. The energy $\phi_{LS}$ can be decomposed into two potentials due to a semi-infinite liquid phase: in one case the solid element is at a distance $-z$ from the semi-infinite liquid phase, whereas in the other case it is at a distance $h-z$.}
\label{DecompositionC}
\end{figure}
\begin{figure}[t!]
\includegraphics{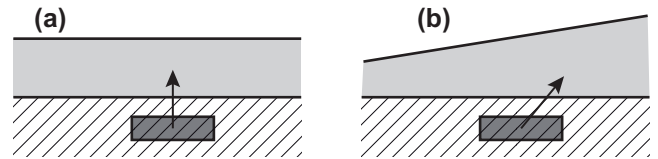}
\caption{(a) For a flat film the force exerted by the liquid on a volume element of solid is in the direction normal to the interface. (b) When the liquid-vapour interface is not parallel to the substrate, the left-right symmetry is broken  a tangential component appears. The bias is towards the side where the amount of attracting liquid is greater. For a postive slope $dh/dx$, this is oriented in the positive $x$ direction.}
\label{Interface_shape}
\end{figure}

\subsection{Tangential capillary force near a contact line}

To identify the capillary forces near a contact line we now consider a microscopic liquid film. We first derive the component of the liquid-on-solid force that is tangential to a flat solid interface. From (\ref{eq:force}) one can infer that it results from the long-range term involving $\phi_{LS}$, since $p_r$ acts along the direction normal to the surface. Suppose that the contact line extends in the $y$-direction, so that the liquid-vapor interface can be written as $h(x)$. Since $\partial /\partial y=0$, the tangential force $F_t$ is along the $x$-direction and can be expressed as 
\begin{equation}
F_t =-\int dy \int_{-\infty}^{\infty}dx\int_{-\infty}^{0}dz\frac{\partial}{\partial x}\left[\phi_{LS}\right],
\label{tang_force}
\end{equation}
where the integrals run over the entire solid domain. 

We now apply the local approximation to determine $\phi_{LS}$. At a position $z < 0$ inside the solid, the potential due to the finite slab of thickness can be expressed as (Fig.~\ref{DecompositionC}):
\begin{equation}
\phi_{LS} = \Pi_{SL}\left(-z \right)-\Pi_{SL}\left(h\left(x\right)-z\right).
\label{diff_inft_domian}
\end{equation}
Here we used the symmetry property $\Pi_{\alpha\beta}=\Pi_{\beta\alpha}$.
Consequently,
\begin{equation}
-\frac{\partial}{\partial x}\left[\phi_{LS}\right]=h^\prime \, \frac{\partial}{\partial h}\left[\Pi_{SL}\left(h-z\right)\right],
\label{deriv_phiLS}
\end{equation}
where $h^\prime=dh/dx$. This implies that there will be a finite contribution of the tangential force only when the liquid film thickness varies with $x$ ($h^\prime\neq0$). This breaking of left-right symmetry, illustrated in Fig.~\ref{Interface_shape}, induces a tangential force towards the side where the greater amount of liquid is present. Replacing (\ref{deriv_phiLS}) in (\ref{tang_force}), we can integrate out the $z$-direction and write the tangential force as,
\begin{equation}
F_t =-\int dy\int_{-\infty}^{\infty}dx\, h^\prime \Pi_{SL}\left(h\right),\label{red_tang_force}
\end{equation}
where we used the property $\Pi_{SL}(\infty)=0$. 

From this expression one can draw two important conclusions. First, one can interpret the argument of the integral as the tangential traction $f_t$, defined as the tangential force per unit area:
\begin{equation}\label{eq:tractiont}
f_t (x) =- h^\prime \, \Pi_{SL}\left(h(x) \right).
\end{equation}
This tangential capillary traction will be used below as input to compute the elastic deformation. Strictly speaking, this force is not localized at the solid-liquid surface, but it is spread out over a thin region in the $z$-direction. In the elastic calculation we will treat $f_t$ as localized on the surface. Second, the \emph{total} tangential force can be quantified exactly from (\ref{red_tang_force}). Writing $\int dx \,h^\prime=\int dh$, one can use the integral property (\ref{int_phi_SL}), to find the central result $F_t/L= \gamma(1+\cos \theta)$ (with $\int dy=L$), already presented in Eq.~(\ref{eq:Ft}). 

Actually, Eq. (\ref{eq:Ft}) does not rely on the local approximation and can be derived exactly,  for arbitrary values of the contact angle $\theta$. Starting again from Eq. (\ref{tang_force}),  the integration over $x$ leads to:
\begin{equation}
F_t=-\int dy\int_{-\infty}^0dz\left(\left.\phi_{LS}\right|_{x=\infty}-\left.\phi_{LS}\right|_{x=-\infty}\right)
\label{def_F_t_1}
\end{equation}
We assume that the liquid bulk is located in the positive $x$ direction, giving $\phi_{LS}|_{x=\infty}=\Pi_{LS}(-z)$. At the same time, the dry solid or the precursor film is located in the negative $x$ direction, $h(-\infty)=h_0$. Using the construction of Fig.~\ref{DecompositionC} with film thickness $h_0$, this gives $\phi_{LS}|_{x=-\infty}=\Pi_{LS}(-z)-\Pi_{LS}(h_0-z)$. 
Inserting these expressions in (\ref{def_F_t_1}), we get
\begin{eqnarray}
F_t/L&=&-\int_{-\infty}^0 dz \,\Pi_{SL}\left(h_0-z\right) \nonumber \\
&=&-\int_{h_0}^\infty d\tilde{z} \,\Pi_{SL}\left(\tilde{z}\right)
=\gamma(1+\cos{\theta}),
\label{def_F_t_2}
\end{eqnarray}
where we once more used the normalization (\ref{int_phi_SL}). 

\subsection{Normal capillary force near a contact line}

\begin{figure}[t!]
\includegraphics{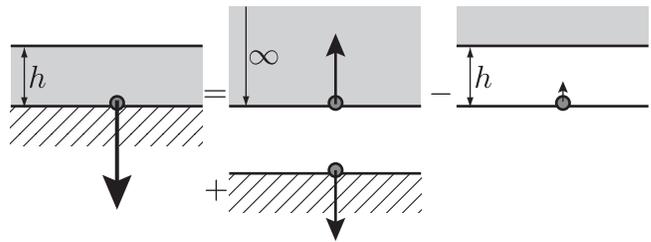}
\caption{The interaction of a volume element of liquid, located at the solid-liquid interface, can be decomposed into the interaction exerted by the solid $\phi_{SL}$, plus two interactions between the same liquid element with a semi-infinite liquid phase (in one case at a distance $h$ and the other case at zero distance).}
\label{DecompositionB}
\end{figure}

The normal force near a contact line can again be computed from (\ref{eq:normal}), but now we take into account that the liquid thickness is finite and lies within the range of molecular interaction.   For a liquid profile $h(x)$, the expression for the normal force becomes
\begin{equation}
F_n= - \int dy \int_{-\infty}^{\infty}dx \left( p_r  + \phi_{LS} \right)_{z=0}.
\end{equation}
From Fig.~\ref{DecompositionC}, we compute the potential inside the solid, located at the solid-liquid interface as 
\begin{equation}
\phi_{LS}=\Pi_{LS}(0) - \Pi_{LS}(h)=\Pi_{SL}(0) - \Pi_{SL}(h)\;.
 \end{equation}
To determine $p_r$ we again use the equilibrium condition (\ref{eq:P}) inside the liquid, and evaluate it at the solid-liquid interface. This requires the potentials $\phi_{LL}$ and $\phi_{SL}$  at the solid-liquid interface, as sketched in Fig.~\ref{DecompositionB}. The potential due to the solid is simply $\phi_{SL} = \Pi_{SL}(0)$, since the solid is an infinite half space. Using the local approximation, the liquid-liquid interaction can be expressed in terms of $\Pi_{LL}$ (cf. Fig.~\ref{DecompositionB}): 
\begin{equation}
\phi_{LL} = \Pi_{LL}(0) - \Pi_{LL}(h)\;.
\end{equation}
Hence, one obtains from (\ref{eq:P}):
\begin{equation}\label{eq:P2}
p_r = P_L+ \Pi_{LL}(h) -\Pi_{SL}(0).
\end{equation}
This gives the final expression for the normal force:
\begin{eqnarray}
F_n &=& - \int dy \int_{-\infty}^{\infty}dx \left\{P_L+  \Pi_{LL} (h) - \Pi_{SL}(h) \right\}\nonumber \\
&=& - \int dy \int_{-\infty}^{\infty}dx \left\{  P_L-\Pi(h) \right\} . \label{eq:finalnormal}
\end{eqnarray}

This result can be interpreted as follows. The argument of the integral represents the normal traction $f_n$ that can be used to compute the elastic deformation:
\begin{equation}\label{eq:tractionn}
f_n (x) =-P_L+ \Pi\left(h(x) \right).
\end{equation}
This is indeed identical to the normal traction proposed by White~\cite{WhiteJCIS}. The disjoining pressure contribution $\Pi(h)$ acts only in the vicinity of a contact line, where $h$ is in the range of the molecular interactions. It is responsible for the ``pulling" force at the contact line. The second term, $P_L$, is simply the thermodynamic pressure in the liquid, which was already identified in Sec.~\ref{sec:laplace} for a macroscopic drop. 

To quantify the magnitude of the pulling force, we now consider a two-dimensional profile for an infinite drop with zero external pressure, i.e. $P_L=P_v=0$. Some numerical examples will be presented in the next section (cf. Fig.~\ref{2D_drop_fig}). The equilibrium profile is then given by a balance between Laplace pressure and disjoining pressure, which according to (\ref{eq:dropeq}) can be written as
\begin{equation}
\Pi\left(h\right)+\gamma\kappa=\Pi\left(h\right)-\gamma\frac{h^{\prime\prime}}{\left(1+h^{\prime 2}\right)^{3/2}}=0.
\label{eq_eqn_2D}
\end{equation}
This can be integrated to~\cite{WhiteJCIS}
\begin{eqnarray}
\gamma \left(\frac{h'}{(1+h'^2)^{1/2}}\right)_{x=\infty} &=& \int_{-\infty}^\infty dx \, \Pi(h),
\end{eqnarray}
where we used $h'(-\infty)=0$ corresponding to the dry solid or the precursor film. Thus with $h^\prime(\infty) = \tan \theta$, the left-hand side becomes $\gamma \sin \theta$. The right-hand side is an integral over the normal traction $f_n$. Hence, the integrated normal force at the contact line becomes, for $P_L=0$,
\begin{equation}
F_n =  \int dy 
\int_{-\infty}^\infty dx \, \Pi (h(x)) 
= \int dy \, \gamma \sin \theta,
\end{equation}
which is the anticipated result (\ref{eq:Fn}).

\section{Numerical results} \label{sec:numerics}

\subsection{Drop shape and capillary traction}

From the analysis above it is evident that in order to compute the spatial structure of tangential and normal tractions,  one needs to know the exact shape of the drop. This can be done once the disjoining pressure has been specified. As an illustration of our results we consider a disjoining pressure $\Pi(h)$ that corresponds to a Lennard-Jones (9-3) potential~\cite{SellierBMF, IngebrigtsenJPCC},
\begin{eqnarray}
\Pi_{SL}\left(h\right)&=&\frac{A}{h_0}\left[\left(\frac{h_0}{h}\right)^9-\left(\frac{h_0}{h}\right)^3\right],
\\
\Pi_{LL}\left(h\right)&=&\frac{B}{h_0}\left[\left(\frac{h_0}{h}\right)^9-\left(\frac{h_0}{h}\right)^3\right],
\\
\Pi\left(h\right)&=&\frac{B-A}{h_0}\left[\left(\frac{h_0}{h}\right)^9-\left(\frac{h_0}{h}\right)^3\right],
\label{Disj_Press_specf}
\end{eqnarray}
without any correction due to the pair correlation function. The molecular length $h_0$ is the equilibrium thickness of the precursor film and sets the length scale for the interactions. For simplicity we take the same functional form for liquid-liquid and solid-liquid interactions, but this is not required. The second term is the familiar long-range van der Waals attraction, while the first represents the short-range repulsion. Such a repulsion is needed to prevent matter from collapsing and ensures that the surface tensions remain finite, cf. (\ref{int_phi_LL},\ref{int_phi_SL},\ref{int_phi_Disj_h}). In this case, the repulsive term balances the attraction at a molecular length $h_0$, which selects the equilibrium thickness of a precursor film. To illustrate the robustness of our results, Appendix~\ref{Ap_2} also discusses a disjoining pressure for which the repulsion is modeled by a vanishing pair correlation function. Despite the fact this model does not lead to the formation of a precursor film, the resulting tractions on the solid turn out very similar.

The constants $A$ and $B$ are determined by the integral properties (\ref{int_phi_LL},\ref{int_phi_SL},\ref{int_phi_Disj_h}), as $A=\frac{8}{3}\gamma\left(1+\cos{\theta}\right)$ and $B=\frac{16}{3}\gamma$. Below we compute the shape of the drop and the corresponding capillary traction $f_{n,t}$ for two cases: an infinite two-dimensional drop and a finite axisymmetric drop.

\subsection{Infinite two-dimensional drop}
The drop shape for an infinite two-dimensional drop can be determined by the equilibrium condition (\ref{eq_eqn_2D}). The boundary condition used for the solution of the drop profile is that the interface asymptotically joins the precursor film, $h(-\infty)=h_0$. This automatically gives the correct contact angle as $h'(\infty)=\tan \theta$. 

\begin{figure}[ht]
\includegraphics{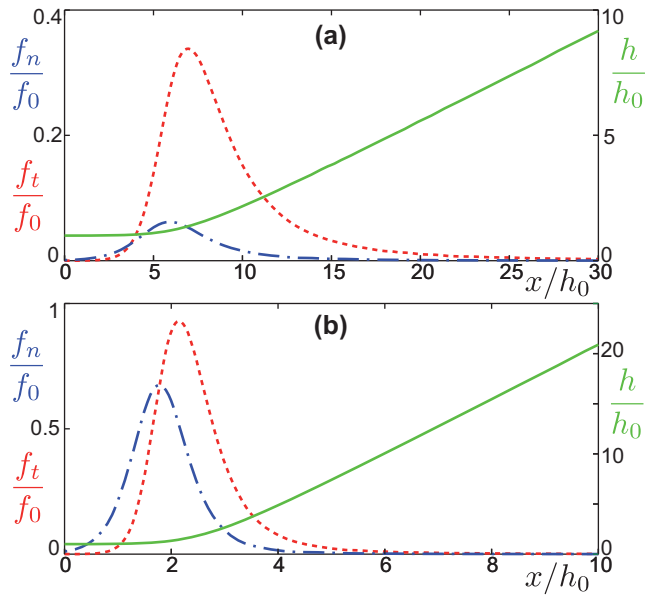}
\caption{(Color Online) Drop shape and capillary tractions for an infinite drop with (a) $\theta=20^{\circ}$ and (b) $\theta=70^{\circ}$. In both plots the drop profile $h(x)/h_0$ is represented by a solid line (green), the tangential traction $f_t/f_0$ by a dashed line (red) and the normal traction $f_n/f_0$ by a dash-dotted line (blue). Tractions are normalized by the pressure scale $f_0=\gamma/h_0$.} 
\label{2D_drop_fig}
\end{figure}

Figure~\ref{2D_drop_fig} shows the drop shape and the capillary traction for two contact angles, $\theta=20^\circ$ and $70^\circ$. The solid line (green) represents the shape of the liquid $h(x)$ normalized by the precursor film thickness $h_0$. The peaked curves are the traction $f_n$ (dash-dotted line in blue) and $f_t$ (dashed line in red), normalized by the natural pressure scale in the model $f_0 = \gamma/h_0$. Indeed, one can see that the capillary forces are localized in the vicinity of the contact line, in between the precursor film and the macroscopic drop. The width of the peaks is set by the length scale $h_0$. The most important result highlighted in these figures is that for a small contact angle, the tangential traction is significantly larger than the normal traction -- this is consistent with Eqs.~(\ref{eq:Fn},\ref{eq:Ft}) since the total force (per unit length) is obtained by integrating over the peaks. In fact, this integral property is determined by the values of the surface tensions and holds irrespective of the choice of disjoining pressure. Only the detailed shape of the peaks is affected by the choice of disjoining pressure -- compare e.g. to the example given in Appendix~\ref{Ap_2}. It is worth noting that, although the local approximation may not be fully quantitative, the figure is still representative since the localization of the capillary forces and their relative strengths are independent of this approximation.

\subsection{Axisymmetric drop}

For three-dimensional axisymmetric drops, the equilibrium equation (\ref{eq:dropeq}) for the drop shape $h(r)$ reads~\cite{MilkavcicJPA}
\begin{equation}
\gamma \kappa +\Pi\left(h\right) = P_L-P_v,
\label{eq_eqn_axsm}
\end{equation}
where the expression of the curvature $\kappa$ is now
\begin{equation}
\kappa = - \frac{h^{\prime\prime}}{\left(1+h^{\prime 2}\right)^{3/2}} - \frac{h^{\prime}}{r\sqrt{1+h^{\prime 2}}}.
\label{eq:curvature}
\end{equation}
This equation is solved numerically with the boundary conditions $\left(h^\prime\right)_{r=0}=0$ (symmetry) and $\left(h^\prime\right)_{r\rightarrow\infty}\rightarrow 0$ (matching onto a precursor film). The same Lennard-Jones disjoining pressure is used. Different drop volumes can be obtained by varying $P_L-P_v$.  

\begin{figure}[h]
\includegraphics{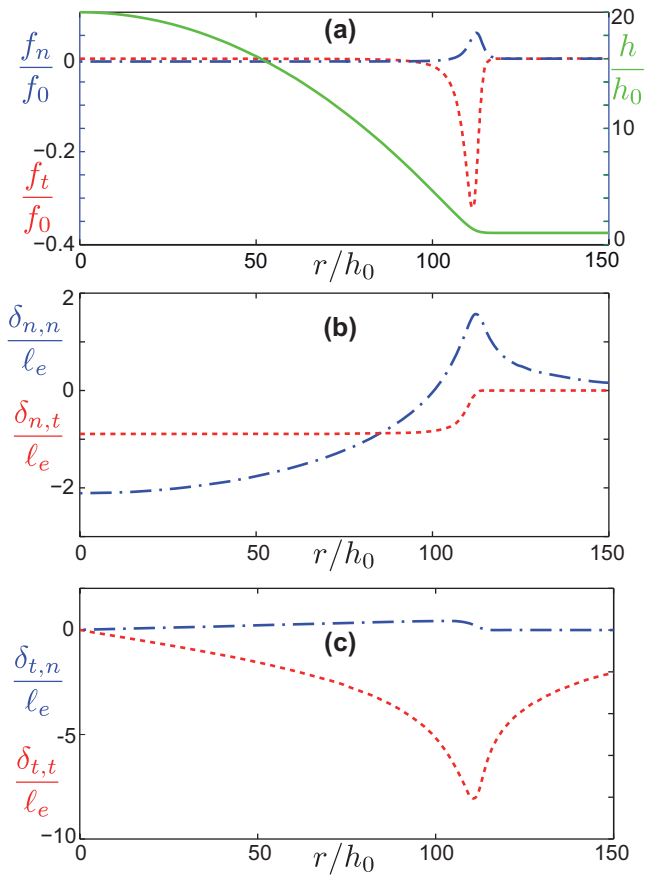}
\caption{(Color Online) Radial profiles of height, stresses and displacements for an axisymmetric drop. (a) Drop profle (solid green line); tangential traction $f_t$ (dashed red line); normal traction $f_n$ (dashed-dotted blue line). (b) Normal deformation due to normal (dashed-dotted blue line) and tangential (dashed red line) tractions. (c) Tangential deformations due to normal (dashed-dotted blue line) and tangential (dashed red line) tractions.}
\label{Axsm_drop_fig}
\end{figure}

Figure~\ref{Axsm_drop_fig}a shows the drop shape $h(r)$ and normalized tractions $f_t$ and $f_n$ for a contact angle $\theta=20^\circ$. Similar results are obtained for different volumes and contact angles. The thickness of the precursor film is slightly different from $h_0$ due to the finite volume, and is set by $\Pi\left(h\right)=P_L-P_v$. Once more the tractions are strongly peaked near the contact line. The negative sign for $f_t$ signals that it acts in the negative radial direction, i.e. towards the centre of the drop. A key difference with respect to infinite drop of Fig.~\ref{2D_drop_fig} is that the normal traction attains a nonzero constant value at the interior of the drop. This is simply the Laplace pressure (equal to $0.0061\gamma/h_0$ here), with a minus sign. 

This figure forms the quantitative basis for ``cartoon" of the capillary forces that was presented in Fig.~\ref{fig:sketch}. From the sign of the normal traction, it is clear that inside the drop the solid is pushed vertically downwards by the Laplace pressure, whereas at the contact line it is pulled upwards by a force that decays rapidly on either sides of the contact line. Similarly, there is a localized tangential force pointing towards the interior of the drop. 

\section{Elastic deformations below a liquid drop}\label{sec:elastic}

With the drop shapes and the tractions on the solid in hand, one can compute the elastic response of the solid substrate. We will use the tractions $f_t$ and $f_n$ from Eqs.~(\ref{eq:tractiont}) and (\ref{eq:tractionn}), as obtained for the undeformed solid. Hence, we assume that the deformation of the solid does not affect the resulting capillary traction, which in practice means $\gamma/h_0\ll E$. In addition, we treat the tractions as perfectly localized at the interface. Ideally, one would like to treat the solid as a semi-infinite elastic body and consider the tractions as a simple load per unit length. This cannot be done, however, since such a line-loading gives rise to logarithmically diverging displacements in the far field~\cite{Johnson}. This problem can be avoided by considering a finite amount of liquid, which is commonly done using axisymmetric drops of finite volume~\cite{BasaPOF}. An alternative is to consider an elastic medium that has finite thickness~\cite{YuJCIS}, which is mathematically much more involved. As already mentioned, none of this literature considers the elastic deformations due to tangential capillary traction. 

Here we consider elastic deformations induced by axisymmetric drops, whose tractions are functions of $r$ only. The elastic response then follows from standard expressions for semi-infinite elastic bodies that can be found e.g. in  \cite{Johnson}. We distinguish the following cases:
\begin{itemize}
\item{normal displacement due to normal traction ($\delta_{n,n}$)}
\item{normal displacement due to tangential traction ($\delta_{n,t}$)}
\item{tangential displacement due to normal traction ($\delta_{t,n}$)}
\item{tangential displacement due to tangential traction ($\delta_{t,t}$)}
\end{itemize}

The normal displacement at the solid surface due to normal traction can be obtained as follows:
\begin{equation}
\delta_{n,n}\left(r\right)=\frac{1-\nu^2}{\pi E}\int_0^{\infty}\frac{4r^\prime}{r+r^\prime}K\left(m\right)f_n\left(r^\prime\right)dr^\prime 
\end{equation}
Here $\nu$ is the Poisson ratio, $E$ is the Young's modulus of the material and $K\left(m\right)$ is the elliptic integral of the first kind with $m=\frac{4rr^\prime}{\left(r+r^\prime\right)^2}$. Similarly, the tangential displacement due to tangential traction can be computed as 
\begin{eqnarray}\nonumber
{\delta}_{t,t}\left(r\right)=&&\frac{1-\nu^2}{\pi E}\int_0^a\frac{4r^\prime}{r+r^\prime}f_t\left(r^\prime\right)\times \\
&&\left[\left(\frac{2}{m}-1\right)K\left(m\right)-\frac{2}{m}Y\left(m\right)\right]dr^\prime.
\end{eqnarray}
where $Y(m)$ is the elliptic integral of the second kind. It should be noted that the Kernels display a logarithmic singularity at $r=r'$, i.e. $m=1$. This means that approximating the tractions as Dirac $\delta$ functions perfectly localized at the contact line would give a diverging deformation at the contact line. Instead, with the tractions spread out over a finite width of order $h_0$, the maximum displacement scales as $\sim \ln(R/h_0)$, where $R$ is the drop size. 

By contrast the normal displacement due to tangential forces, $\delta_{n,t}$, can be obtained in a much simpler form, without invoking the elliptic integrals:
\begin{equation}
{\delta}_{n,t}\left(r\right)=-\frac{\left(1-2\nu\right)\left(1+\nu\right)}{\pi E}\int_r^{\infty}f_t\left(r^\prime\right)dr^\prime,
\end{equation}
\newline
and similarly, the deformation $\delta_{t,n}$ follows as 
\begin{equation}
{\delta}_{t,n}\left(r\right)=-\frac{\left(1-2\nu\right)\left(1+\nu\right)}{E}\int_0^{r}\frac{r}{r^\prime} f_n\left(r^\prime\right)dr^\prime.
\end{equation}
These expressions no longer display a singular Kernel. Note that the prefactors of these two expressions vanish for $\nu=1/2$, which corresponds to incompressible elastic materials.

From these expressions one can identify a typical length scale $\ell_e$ for the elastic deformations. Since integrals over $f_n$ and $f_t$ are proportional to the surface tension $\gamma$, we define the elastic length scale as:
\begin{equation}
\ell_e=\frac{\left(1-\nu^2\right)\gamma}{\pi E}.
\end{equation}
Based on this length scale, we can now tabulate the scaling of the deformations in different directions (Table~\ref{tab_elastic}). In these expressions we actually account for the $\theta$ dependencies of the normal/tangential tractions, according to Eqs.~(\ref{eq:Fn}) and (\ref{eq:Ft}). This clearly shows the importance of the tangential capillary forces. For normal displacements, the effect of the tangential traction is comparable to that of the normal traction, in particular when the contact angle $\theta$ is small. For tangential displacements, however, tangential forces even dominate the effect of normal forces and omitting them induces a significant quantitative error. 
\begin{table}[h] 
\caption{Scaling of the displacements in normal and tangential directions due to normal and tangential tractions, expressed in the elastic length $\ell_e=\frac{\left(1-\nu^2\right)\gamma}{\pi E}$.} 
\centering \begin{tabular}{c c c} 
\hline  & $\delta_n/\ell_e $ & $\delta_t/\ell_e$ \\ [0.5ex]	 \hline Normal traction &$\sin{\theta}$&$\frac{1-2\nu}{1-\nu}\pi\sin{\theta}$ \\  \hline Tangential traction &  $\frac{1-2\nu}{1-\nu}\left(1+\cos{\theta}\right)$&$1+\cos{\theta}$ \\ [1ex] \hline \end{tabular} \label{table:nonlin} \label{tab_elastic} 
\end{table}

To further illustrate this, we have computed the deformation profiles for $\theta=20^\circ$, $\nu=0.35$, and for the disjoining pressure given by Eq.~(\ref{Disj_Press_specf}). For the numerical evaluations of the elliptic integrals (necessary for $\delta_{n,n}$ and $\delta_{t,t}$), we used a method discussed in Johnson \cite{Johnson} that explicitly avoids the singular Kernel. The resulting displacement profiles are shown in Figs.~\ref{Axsm_drop_fig}bc. Normal displacement produced by tangential traction is comparable to that produced by the normal traction, whereas the tangential deformation produced by the tangential traction is significantly larger than the corresponding contribution of the normal traction. The displacements $\delta_{n,n}$ and $\delta_{t,t}$ display a maximum near the contact line, whose magnitude depends on the drop size as $\sim \ln(R/h_0)$. For micron-sized or larger drops, the maximum will thus be much more pronounced than shown here since $h_0$ is of molecular scale. A peculiarity of $\delta_{n,t}$ and $\delta_{t,n}$ is that the displacements are nonzero only at the interior of the drop, and can thus not be measured outside the drop.

\section{Summary and perspective}

In this paper, we have used a nanoscopic yet continuum approach --~the Density Functional Theory in the sharp kink approximation~-- to derive the capillary forces exerted on a solid substrate by a liquid drop resting on it. The drop exerts a large tangential capillary force in the vicinity of the contact line, with a resultant magnitude of $\gamma\left(1+\cos{\theta}\right)$ per unit length. Such a force has not been reported before and its physical origin is sketched in Fig.~\ref{fig:sketch}c: there is a resultant attraction towards the interior of the drop, where most of the liquid is situated. Clearly, this bias is of geometric origin and therefore does not rely on details of the sharp kink approximation. We thus expect that a more complete theory only gives minor quantitative differences. Let us note that our model recovers the normal component of the capillary force, which is known to have a resultant $\gamma\sin{\theta}$. In the bulk of the drop, the Laplace pressure exerts a vertical push on the substrate. 

This novel picture of capillary forces is next employed to determine the elastic deformations of the solid. This is relevant when the substrate is sufficiently soft --~i.e. when the elastic modulus is sufficiently small~-- to lead to nanoscopic or microscopic displacements. The principal limitation of our elastic calculation is that it only applies for small deformations, i.e., $\gamma/h_0\ll E$. At larger deformation, the resulting change in geometry of the solid will have a significant effect on the capillary tractions, as demonstrated in the recent work by Jerison et al~\cite{JerisonPRL}.  Another approximation is that we treated the tractions as localized at the surface of the substrate. However, the corrections induced by the finite range of the interactions will have an effect only in a small region around the contact line, where e.g. the maximum displacements are  known to depend logarithmically on details of the regularization~\cite{Johnson,RusanovUSSR, WhiteJCIS}. 


The predicted tangential force is substantial and will have a measurable effect on the elastic deformations. Typically, experiments studying the solid deformation below a liquid drop measure the \emph{normal} displacements~\cite{PericetLANGMUIR,CamaraCPC,CamaraSM} -- see e.g. Fig.~\ref{fig:sketch}a. This is equivalent to $\delta_{n,n}+\delta_{n,t}$ in the present analysis. These substrates often have a Poisson ratio close to $\nu= 0.5$, in which case the contribution due to the tangential force $\delta_{n,t}$ is expected to be relatively small. By contrast, the tangential deformations of the free surface are very sensitive to the tangential force, as can be inferred from Fig.~\ref{Axsm_drop_fig} and Table~\ref{tab_elastic}. We thus suggest that, in principle, tangential displacements should provide a good experimental tool to access tangential capillary forces. 


Finally, let us emphasize that a tangential force below the contact line of magnitude $\gamma\left(1+\cos{\theta}\right)$ provides a striking perspective on capillary forces. We first note that this does not lead to a violation of YoungÕs law for the contact angle: the computed tractions are exerted on the solid, while YoungÕs law reflects the forces exerted on the liquid. However, when projecting the surface tension along the liquid-vapor interface, the tangential component would be $\gamma \cos{\theta}$~\cite{LiuPHYSICAB} and not $\gamma\left(1+\cos{\theta}\right)$. Similarly, when considering a plate plunged in a liquid bath, the virtual work principle dictates that the resultant force on the solid in fact is $\gamma \cos{\theta}$~\cite{deGennes}. All these thermodynamic results are correct and do not contradict our present analysis. Namely, $\gamma\left(1+\cos{\theta}\right)$ is the force per unit length exerted \emph{below the contact line} -- there are other forces exerted on the substrate at other places. For instance, the tangential forces around a drop balance each other. The case of the plunging plate is more subtle. We show in the Appendix~\ref{ap_1}F that a normal force appears on the solid in places where the substrate is curved. For the plunging plate, there are thus very strong forces acting on the bottom edges of the plate. Indeed, one can show that these ensure that the overall force on the solid is in accordance with the virtual work principle. Once again, we thus expect direct experimental evidence of  the tangential force $\gamma\left(1+\cos{\theta}\right)$, by measuring the tangential compression of a soft elastic plate in a liquid.

{\bf Acknowledgements:~}We gratefully acknowledge discussions with Joost Weijs and Kees Venner.

\appendix

\section{Tractions for a different interaction potential}\label{Ap_2}

In this appendix we verify that the tractions shown in Fig.~\ref{2D_drop_fig} are representative in the sense that results are only mildly affected by the choice of disjoining pressure. In particular, we will use an interaction model that does not lead to a precursor film, but has a well-defined contact line at which $h=0$. For this, we consider an interaction
 
\begin{equation}
\varphi_{\alpha\beta}(r)=-\frac{c_{\alpha\beta}}{\left(h_0^2+r^2\right)^3},
\label{eq:int_pot}
\end{equation}
representing a van der Waals attraction that is regularized at molecular scale $h_0$. The regularization stems from the vanishing pair-correlation function within a distance $h_0$. For numerical convenience we follow~\cite{GettaPRE,WeijsPF} and use a simplified form that is easily integrated. From this potential we can derive the disjoining pressure from~(\ref{eq:gen_Pi_phi_rel}), 

\begin{equation}
\Pi_{\alpha\beta}(h)=-\frac{\pi c_{\alpha\beta}}{4h_0^3}\left[\frac{\pi}{2}-\tan^{-1}\left(\frac{h}{h_0}\right)-\frac{h/h_0}{1+(h/h_0)^2}\right].
\label{eq:Pi_new}
\end{equation}
Employing (\ref{eq:Pi_new}) in (\ref{int_phi_LL}) and (\ref{int_phi_SL}), we get

\begin{eqnarray}
c_{LL}&=&\frac{8h_0^2\gamma}{\pi},
\label{eq:cLL}\\
c_{SL}&=&\frac{4h_0^2\gamma(1+\cos{\theta})}{\pi},
\label{eq:cSL}
\end{eqnarray}
and hence the total disjoining pressure becomes
\begin{equation}
\Pi(h)=\frac{\gamma(1-\cos{\theta})}{h_0}\left[\frac{\pi}{2}-\tan^{-1}\left(\frac{h}{h_0}\right)-\frac{h/h_0}{1+(h/h_0)^2}\right].
\label{eq:Pi_disj_new}
\end{equation}

\begin{figure}[t!]
\includegraphics{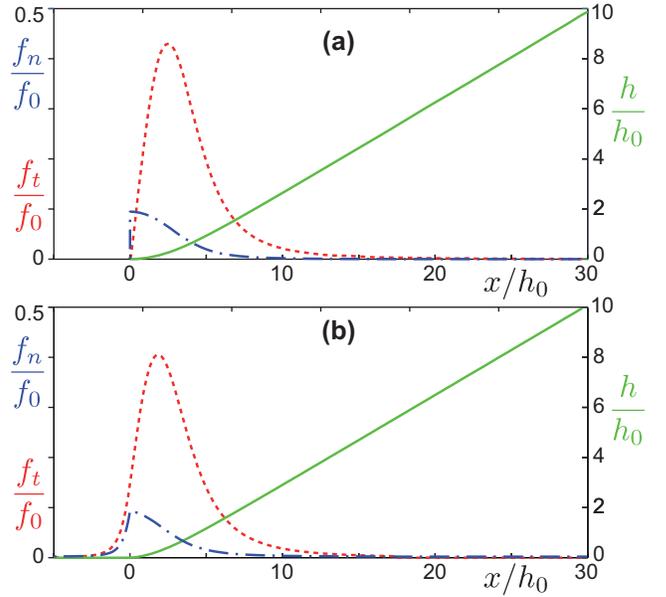}
\caption{Drop shape and capillary tractions for a two-dimensional infinite drop for a model that does not lead to a precursor film. Molecular interactions given by (\ref{eq:int_pot}), contact angle $\theta=20^\circ$. (a) Numerical results obtained from the ``local approximation''. (b) Numerical results obtained from the full DFT model, without invoking the local approximation. See text for details. Overall, the results are very similar to those of Fig.~\ref{2D_drop_fig}a. Note that the local approximation induces a discontinuity in the normal traction, which is not present in the full model.}
\label{Res_new_intr}
\end{figure}

We consider the shape of an infinite two-dimensional drop corresponding to this disjoining pressure. For this we numerically evaluate (\ref{eq_eqn_2D}) with the equilibrium boundary conditions at the contact line, namely $h(0)=0$ and $h^\prime(0)=0$. Figure~\ref{Res_new_intr}a shows the resulting drop shape as well as the tractions from (\ref{eq:tractiont}) and (\ref{eq:tractionn}). These results should be compared to Fig.~\ref{2D_drop_fig}a. First, one notes that the new model does not lead to the formation of a precursor film. Yet, the resulting tractions display very similar characteristics as those obtained from the model with precursor film. The total force (per unit length) is obtained by integrating over the peaks and these are once more equal to $\gamma\sin{\theta}$ and $\gamma(1+\cos{\theta})$, for the normal and the tangential components. Only the detailed shape of the peaks is affected by the choice of disjoining pressure, though for each of the cases the peaks have a typical width $\sim h_0$.

A peculiar feature of the model without the precursor film is that the presence of the contact line seems to induce a discontinuity in the tractions. Namely, outside the drop, $x<0$, the tractions are strictly zero. Inside the drop, the normal traction is finite since $\lim_{h\rightarrow 0^+}\Pi(h)=\Pi(0)\neq 0$, and hence leads to a discontinuity. We show, however, that this discontinuity is an artifact of the local approximation of the DFT model. This approximation implicitly assumes small spatial derivatives ``$d/dx$", a condition that is clearly violated near the contact line in case there is no precursor film. Using the methods described in~\cite{WeijsPF}, we were able to numerically solve the full DFT model for the interaction~(\ref{eq:int_pot}). By this we mean that we solve the drop shape and tractions from the functionals described in Sec.~\ref{sec:forces}, without invoking the local approximation. The results are presented in Fig.~\ref{Res_new_intr}b. Indeed, the tractions are very similar to those predicted by the local approximation, the main difference being a smoothening of the discontinuity near $x=0$.

\section{Curvature induced force} \label{ap_1}

\begin{figure}[h!]
\begin{center}
\includegraphics{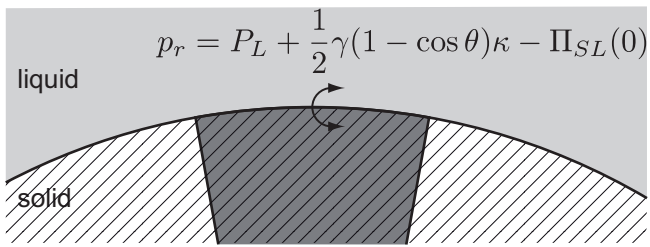}
\caption{\label{CurvedSolid}Repulsion $p_r$ at the surface of a curved solid. By convention, the curvature $\kappa$ is positive on the schematic.}
\end{center}
\end{figure}

In this appendix, we compute the force exerted by the liquid on a solid of curvature $\kappa$, depicted in Fig.~\ref{CurvedSolid}. From the point of view of the liquid, the same interface has thus a curvature $-\kappa$. In this case the expression of the normal stress on the solid is slightly modified with respect to the flat surface, cf Eq.~(\ref{eq:normal}), as it picks up a curvature contribution:
\begin{eqnarray}
\label{curvedstress}
f_n&=&- p_r  - \phi_{LS} + \kappa \int_{0}^\infty\Pi_{SL}\left(h\right)dh \nonumber\\
&=&- p_r - \phi_{LS} - \gamma(1+\cos\theta)\kappa \;.
\end{eqnarray}
Considering the sub-system shown in dark gray in Fig.~\ref{CurvedSolid} the curvature correction term comes from the reduced volume that is attracted to the liquid, as compared to the flat case. If the solid is convex, i.e. $\kappa>0$, this reduction of attraction effectively yields an increased pressure.

As previously,  repulsion can be expressed using the total potential. Using the analogue of equation (\ref{laplaceLL}) for the solid-liquid interaction, one express the potentials:
\begin{eqnarray}
\phi_{LL}&=&\Pi_{LL}(0)-\gamma\;\kappa\;,\\
\label{laplaceSL}
\phi_{SL}&=&\Pi_{SL}(0)+ \frac{1}{2}\gamma \left(1+\cos\theta\right) \kappa\;,\\
\label{laplaceLS}
\phi_{LS}&=&\Pi_{LS}(0)-\frac{1}{2}\gamma \left(1+\cos\theta\right) \kappa\;.
\end{eqnarray}
By  symmetry, we also have $\Pi_{LS}(0)=\Pi_{LS}(0)$. From Eq.~\ref{eq:P}, we get the expression of the repulsion:
\begin{equation}
p_r = P_L+\frac{1}{2}\gamma (1-\cos \theta)\kappa - \Pi_{SL}(0)
\end{equation}
Finally, we obtain from (\ref{curvedstress}):
\begin{equation}
f_N=-P_L-\gamma\kappa\;.
\end{equation}
Hence, the solid is pushed by the liquid towards the inside of the curvature. Surprisingly, the solid-vapor and solid-liquid surface tensions do not appear in the expression.

\bibliographystyle{jasanum}
\bibliography{ElastoPOF}

\end{document}